\documentstyle [12pt]  {article}
\evensidemargin\oddsidemargin
\begin{document} 
\count0 = 1    
%\begin{titlepage}
\vspace{30mm}
 \title{{ Fuzzy Geometry of Phase Space  \\    
  and Quantization of Massive Fields \\ }}
\author {S.N.Mayburov \\
% \thanks{E-mail ~~ maybur@sgi.lpi.msk.su  ~~
Lebedev Inst. of Physics\\
Leninsky Prospect 53,
Moscow, Russia, RU-117924\\
 E-mail: mayburov@sci.lebedev.ru \\}
\date { }
\maketitle
\begin{abstract}

The quantum space-time
and the phase space  with fuzzy  structure  
is investigated as the possible  quantization formalism.
 In this theory the state of nonrelativistic
 particle $m$  corresponds to the element of
fuzzy ordered set (Foset)  - fuzzy point.
Due to Foset partial (weak) ordering, $m$ space coordinate 
$x$ acquires principal uncertainty $\sigma_x$.
It's shown that Shroedinger formalism of 
Quantum Mechanics can be completely  derived from  
consideration of $m$ evolution in fuzzy phase space
 with minimal number of axioms.  
%  reproduces the main quantum effects, in particular
%  the  interference of quantum states.  

\end{abstract}
%
% \pacs { 03.65.Fd, 02.40. -k,98.80.Jk  }
%   \submitto {\JPG} ???
%\vspace{12mm}
  \small{Talk given at Workshop 'Quantum Field Theory\\
 under Influence of External conditions'\, 
 Leipzig, sept. 2007 \\}
%\end{titlepage}
%
%\maketitle
%
\section { Introduction}

Quantum  space-time  and its  relation to axiomatics of
Quantum Mechanics (QM) and Field Theory is
  actively discussed now under the different angles \cite {Vax,May,Aha}.
% The interest to it enforced
% by the   indications that its structure  at small (Plank) scale
% can be quite nontrivial \cite{Dop,Ish}.
%  Due to the absence of any experimental 
% information it seems instructive to look for some indications 
% reconsidering under this angle  standard Quantum Physics space-time picture.
In particular, it was proposed that
 the  fundamental properties of space-time 
  metrics and topology
 can be modified significantly at Plank scale \cite {Ish,Vol}.
% This properties are found in P-adic QM inspired by Numbers Theory \cite {Vol}.
%
Our work motivated largely by this ideas, 
 which will be studied in the framework of Sets Theory,
 exploring the various set structures  of space-time manifold $M_{ST}$.
 For example, in 1-dimensional Euclidean Geometry, 
 the elements of its manifold $X$  - the points
$\{x_a\}$  constitute the  ordered set.
%  Classical particle $m$ state
% corresponds to $M_{s-t}$ point $x(t)$ and formally this state is the
%  point $r_p(t)$ in phase space.
%  In QM one regards as particle's  state the extended object - Dirac
%  state vector $\Psi$ evolving on the same  $M_{s-t}$ manifold 
%  ,i.e  the classical space-time transferred to QM copiously.
Yet there are  other kinds of fundamental sets  
which also permit to construct the  consistent geometries on them.
In this paper we shall investigate  
  Posets and the fuzzy ordered sets (Fosets); in this case, their elements 
can be incomparable or weakly ordered relative to each other \cite {Zad,Got}.
Basing on Foset structure, 
novel Fuzzy Geometry  was constructed   
which will be studied here as the temptative space-time and
 phase space geometry \cite {Zee,Dod,Pos}.

%  In our approach to quantization  such 
%  constitutes the basic space-time manifold $M_F$
%   with fuzzy relations between its  elements
%  - fuzzy points for which proposition $L_x$ can be untrue
% and only a weak propositions of the kind :'$x_1$ is in $x_2$ vicinity
% of the approximate width $\sigma_x$'characterizes a points relations.

 In  Classical Mechanics in 1-dimensional space
 $X=R^1$ the newtonian particle is defined as  'material'  point $x^m(t)$,
ordered on $R^1$  set, i.e relative to all its elements $\{x_a\}$.
Analogously to it, in our approach
 the massive particle corresponds to the  fuzzy point $b_m(t)$
 in Fuzzy space manifold $C^F$.
%  can be defined on $M^F$ , then  fuzzy points
Due to its weak ordering, such particle possess the principal uncertainty
of $x$ coordinate, i.e. it is smeared 
in $R^1$ space  with an arbitrary dispersion $\sigma_x$ \cite {May}.
%    and in velocity $V_x$ space with $\sigma_v$;
%  and  in momentum $P_x$ space with $\sigma_p$.
In such theory the quantization by itself can be defined
 as the transition from the ordered phase space to fuzzy one, i.e.
the quantum properties of particles and fields are induced directly by
 Fuzzy  Geometry of their  phase space and don't postulated separately
 from its geometric structure.
In this paper as the  simple example of such transition
the  quantization of nonrelativistic particle will be regarded;  
 it will be shown that Fuzzy Geometry 
 induces the particle's dynamics which is equivalent to
Schr$\ddot{o}$dinger QM dynamics.
%
%  Furthermore the particles mechanics  on fuzzy manifold described here
% permit to reproduce the  main quantum effects.

 Considering our approach under more formal angle, it can be regarded as the 
 alternative QM formalism, which probably is more suitable for
  the  quantization in
  arbitrary space-time geometry, which can appear at Plank scale.
 Remind that beside standard Schroedinger
QM  formalism, there several others, which are more or less equivalent, of
 them the most popular are $C^*$-algebras and Quantum Logics \cite {Schiff}.  
Earlier it was shown that  the  fuzzy observables 
are the natural generalization of QM observables \cite {Ali}.
% The  'fuzzy lumps'  were applied in Quantum Gravity
% and Cosmology studies \cite {Ren}. 
 In the last years it was found also
that some  fuzzy sets features
 are appropriate also for Quantum Logics formalism
 (\cite {Pyk} and refs. therein). 
In chap.2 we shall study the structure of particle's states induced by  
Fuzzy Geometry and discuss semiqualitatively the main features of
their evolution.
 Basing on this considerations, in chap.3 the evolution
 equations for free particle and the particle in the external field will 
be derived; it will be shown that they are equivalent to Shr$\ddot o$dinger
 QM formalism. The first results of our theory were published in
\cite {May}.

%  Yet it exploits nonstandard logics
% of propositions with multiple outcomes , besides standard yes/no.
%
%  In distinction we sticks to standard Boolean Logics in our theory
%   and apply in place of it Fuzzy Geometry of phase space. 
% This Fuzzy Geometry  approach to QM can be interpreted
%  as the novel quantization
% formalism i.e.  formal transition from Classical mechanics to QM
% It was formulated already via path integrals, Von Neumann algebra,
% deformations and other approaches.

 \section {Fuzzy Geometry and Fuzzy States}

Now we shall regard the connection between Fuzzy Geometry and  FM,
 analogously to the connection between Euclidean Geometry
 and Classical Mechanics. We shall not review here  fundamentals
 of Fuzzy Geometry, which can be found elsewhere \cite {Dod,Zee}, 
 restricting its consideration only to simple examples important
for our formalism.
% To illustrate  Fuzzy Relations and Fuzzy Geometry features,
% We shall start from consideration of simplest sets,
Remind that for  elements of partial ordered set (Poset)
$D=\{d_i\}$, beside the standard ordering relation between
its elements $d_k\leq d_l$ (or vice versa),
% For example the  element $a_k\in P_A$ ordered relative to some $a_j$
% but isn't ordered relative to some others
the  incomparability relation  $d_k\, \wr \,d_l$ is also permitted;
if it's true then both $d_k \le d_l$ and $d_l \le d_k$ propositions
 are false. To illustrate its meaning, consider
 Poset $D^T=A\cup B$, which  includes  the 
 subset of 'incomparable' elements $B=\{ b_j \} $,
and the ordered subset $A=\{a_i \}$.
 In $A$  the element's    indexes grow
correspondingly to their ordering, so that  $\forall i$,
  $a_i \le a_{i+1}$.
 Any $b_j$ is incomparable at least to one $a_i$.
% The interval $[d_e,d_f]$ is $D^T$ subset, 
% such that its maximal lower (upper) bound  $d_e, d_f \in A$,
% and any $ a_i,\,b_j \in [d_e,d_f]$, if 
%  $d_e\le a_i,\, b_j \le d_f$.
% For the simplicity let's consider $B$ which
% includes only  one element $b_0$. 
Consider some interval $\{a_l,a_{l+n}\}$ , i.e. $D^T$ subset for which
$\forall a_i, b_j; \, a_l \le a_i, b_j \le a_{l+n}; \, n \ge 2$.  
 Let's suppose that  some  $b_j \in \{a_l,a_{l+n}\}$ and  $b_j$ 
is incomparable with all  $\{a_{l},a_{l+n} \}$ internal elements:
  $ b_j \wr a_i;\,$ iff $l+1\le i \le l+n-1 $.
%   i.e. $a_l\leq a_i, a_i\leq a_{l+n}$. $b_0$ ordered to other $a_j \in A^1$
%  and   $a_l \le b_0, b_0 \le a_{l+n}$
In this case,  $b_j$   in some sense is  'smeared' inside $\{a_l,a_{l+n} \}$
interval, i.e. this is the discrete analogue of space coordinate uncertainty,
 if to regard $A$ as the analogue of coordinate axe.
%   Described situation can be interpreted as 
%   $a_0,a_i$ are approximately equal u arbitrary uncertainty  $\sigma_a$.

Fuzzy relations can be regarded as the generalization
of regarded incomparability relations which
 introduces  the positive   measure of incomparability $w$.
To define it, let's put in correspondence
to each $b_j,a_i$ pair of $D^T$ set the weight $w^j_i \ge 0$
 with the  norm $\sum_i w^j_i=1$.
The simplest  example is the homogeneous incomparability:
  $w^j_i=\frac{1}{n}$ for regarded
 $a_i \in [a_l,a_{l+n}]$ interval;
$w^j_i=0$ outside of it. It can be interpreted as $b_j$ homogeneous smearing
inside $[a_l,a_{l+n}]$.
% If $b_0$ is ordered in$A$, for example $b_0=a_i$, then $w^0_j=\delta_{ij}$.  
 If $w$ defined for all $a_i,b_j$ pairs in $D^T$, then
$D^T$ is  Foset $D^F$, and $b_j$ are the fuzzy points \cite {Zee}.
%  Note  that in distinction from 
% the regarded case, in general an arbitrary Foset isn't necessarily Poset. 
%
The continuous 1-dimensional Foset $C^F$ is defined
analogously; $C^F=B \cup X$ where $B$ is the same as above,
% the ordered  subset $A$ is substituted
$X$ is the continuous ordered subset.
 If the constant metrics is defined on
  $X$, then it's equivalent to $R^1$ axe of real numbers.
% In the simplest case one can
% take the same $B$, then $B \in C^T$ is the
% discrete subset of fuzzy points,  we shall regard here only
% $B=\{b_0\}$.
%  The interval $[x_v,x_u]$ is defined analogously to
% the discrete case, 
% if $b_0 \in [x_c,x_d]$ interval, it's    
 Fuzzy relations between $b_j,x_a$  are
described by the continuous distribution
 $w^j(x_a)\ge 0$ with the norm $\int w^j dx=1$, in this case
 $C^F$ is called  the fuzzy space.
% If on $X$ some metrics is defined, $C^F$ is called fuzzy space, below we
% shall regard $C^F$ only as the sum of 1-dimensional Euclidean space and
% the discrete set $B$ of the fuzzy points.
 Note that in Fuzzy Geometry  $w^j(x)$ doesn't have any probabilistic 
(stochastic) meaning but only the algebraic one \cite {Dod}. 
% The fuzzy   structures   to some extent are analogous
% to Orthomodular or BvN algebras  which describes some Quantum 
% Structures  \cite {Pyk}.
% The remarkable analogy between the  uncertainty of 
% fuzzy points  coordinates and QM uncertainties
% was noticed already \cite {Dod}, but to our knowledge
% no consistent development of such ideas was performed.

The particle state in Classical Mechanics corresponds to ordered point
$\{\vec {r}(t),\vec{p}(t)\}$ in 6-dimensional Euclidean phase space $R^3*R^3$.
% We shall consider  the fuzzy point in coordinate space and the  
% description of its evolution by some  state.
% In such classical case the instant  position of the  particle is the point in
%  Euclidean 3-space; its physical state corresponds
%  to the point in the phase or configuration Euclidean 6-space.
%
In FM  the 
nonrelativistic particle $m$ in 1-dimension
 is described as the fuzzy point  $b(t)$ in $C^F$
manifold described above ( its modification for 3-dimensions will be regarded
in  final chapter).
It means that  the particle is  characterized by the positive
density $w(x,t)$  in 1-dimensional space $R^1$ with
 constant norm: $\int wdx=1$.
 It doesn't exclude,
naturally, the existence of other $m$ degrees of freedom
 of which $m$ evolution can depend. 
% below  they will be related to 
%  such $m$ observables as its  momentum $p$  and velocity $v_x$.
In nonrelativistic case, the time $t$ is taken to be the real
parameter on $T$ axe, the particle's evolution
 in FM is assumed to be reversible. FM supposedly
    possess the invariance relative to 
the space and time shifts, also it is invariant
under the space and time reflections.

We concede that $m$  
properties   in  arbitrary  reference frame (RF)
  described by a fuzzy state $ |g (t)\}$, 
  the used notation stresses its difference from
Dirac quantum state $|\psi \rangle$.
In regarded approach, it's natural to 
    start  from assuming that, beside $w(x)$,  
other $ g$ independent components are the real functions of one or
more coordinates, i.e. are the fields:
$$
 \{g^1_{i} (x)\};\, i=1,l_1; \quad
  \{g^2_{j}(x, x')\};\quad j=1,l_2,...\, ,etc.; 
$$
% in the simplest case
% $\breve{g}=\{g_{\mu}(x)\}$ can be the vector field. 
where $g^1_1(x)=w(x)$;
 here and below $t$ is omitted wherever the dependence on it
 is obvious.
The structure  of $|g\}$ states set  $M_s$ isn't postulated, in particular,
it  doesn't assumed to be
 the linear space of any kind  $a\, priory$.
In this  framework, $g$ evolution is supposedly
 described by the first-order
on time differential equation,  it   expressed
by the 'fuzzy' map $|g(t)\}=\hat{U}(t)|g_0\}$ which will be studied
in the next chapter. 
We shall construct FM as the minimal theory i. e. at every stage of its
 formulation it assumed that the number of $|g\}$ degrees of freedom and 
theory free parameters is as minimal as necessary for the theory 
consistency. In general,  FM  formalism will be  
 based  mainly on geometric premises, in the aspect
it's to some extent analogous  to formalism of General Relativity.
 In this chapter we shall try to
find the temptative $|g\}$ structure and
some its evolution properties from simple arguments prompted by 
Fuzzy Geometry.

% In this approach,
% $w(x,t)= F_N(g)$ is  some functional of $|g\}$,
% and  $|g\}$  have
%  $w(x)$ alone can't describe  a future $g$ evolution which can depend on 
%  $w^v(v_x)$, etc.  an
% Beside $w(x)$,  $g$ can include the additional
%  components $\breve g$,  which will be used for the description
% of $m$ evolution parameters;

% Hence we attempt to formulate it as Fuzzy
% Geometrodynamics in which a system evolution defined by the initial 
% geometry.

% The choice of  $g$ components   
%  will be motivated below from the premises of Fuzzy Geometry.

Analogously to QM, besides the pure fuzzy states, we shall use for
 the comparison also
the mixed fuzzy states $g^{mix}$ which are
the probabilistic ensembles of several fuzzy states $|g_i\}$ presented 
with probabilities $P_i$ \cite {Schiff}.
It supposed also that an arbitrary $m$ initial state
$|g_0\}$ can be prepared by some experimental procedure.
To study FM dynamics, it's sensible to start from
the simplest $m$ initial states $|g_0\}$, which are 
point-like with $w_0(x) \sim w^0_1 \delta (x-x_1)$
 or some combinations of
them. In the minimal FM  ansatz for  point-like $m$ initial state (source)
  it results in $m$  density:
$$
       w(x,t)=\Gamma_w (x-x_1,t) w^0_1
$$
where $\Gamma_w$ is  $w$ propagator.
%  $w^O_1 \le 1$, i.e. the
%art of $m$ state can be somewhere far away not influencing $w$.
 The simple example of such evolution 
gives the classical diffusion \cite {Vlad}; in 1-dimension  for 
point-like source in $x=0$ one obtains:
\begin {equation}
  \Gamma_D(x,t)=
     \frac{1}{2\kappa\sqrt{\pi t}}\exp^{-\frac{x^2}{4\kappa^2t}}
                          \label {AAA0}
\end {equation}
where $\kappa$ is the diffusion constant.
%  Naturally, for this source 
%  $w$ dispersion $\sigma_x(t) \to \infty $ at $t \to \infty$.
In this chapter $\Gamma_w=\Gamma_D$ will be used in the toy-model
 illustrating the novel features of FM evolution; the detailed description
of this model can be found in \cite {Vax}. Its exploit is  instructive,
because  the main FM distinction from Classical Mechanics lays in the
correlations between $g$ components in different $x$ points
 and not in the evolution of point-like state.
Moreover, in FM the exact effective  $\Gamma_w$ solutions 
 obtained in the next chapter
 don't differ principally from $\Gamma_D$.
% This solution will be used for the comparison with FM evolution,
% it will help us to reveal the nonclassical FM features.
 
% The measurement of $m$ observables in FM
%   will be discussed in the final part of our
%paper; here we assume only that ${x}$ 
%  $H=\frac{P^2}{2m}+U(x)$.  
%  Separately  should be considered assumption that $|g\}$ 
% has $x$-representation $g(x)$ and $w(x)=F(g(x))$ and thus is local field
%  but we don't assume it  at this stage.
 
% The evolution of any physical object can be described as the map
% of its initial state $g_0$ to the final  $g(t)$;  so
% the evolution of fuzzy point $m$ responds to the fuzzy map
%  = |g(t)\}$. 
%  In Fuzzy Geometry, generally, 
%  $\Xi^f$ maps a fuzzy point $b_j$ or an ordered one $x_i$
%  to some other point $b_k$ or $x_l$ \cite {Dod}.
% Before performing the calculation of $g$ evolution, 
%   $\Xi^f_t$  properties       result in the  important effect of

%To revealit's instructive to start from the study of  simple qualitative
% properties of $\Xi^f_t$ map.
% Of them, the most important

This difference between FM and Classical Mechanics can be illustrated by 
 the effect of $m$ sources smearing (SS) or indistiguishability
which is  the direct analogue
 of quantum  interference. In its essence, depending
on the fuzzy or classical structure
of initial $m$ state (source), $w(x,t)$ form can differ dramatically, 
whereas $\bar x$ will be practically the same.
%  which induces $w(x,t)$ nonlinearity.
%   It constitutes the principal
%   feature of fuzzy evolution and its  origin will be discussed here.
To demonstrate it, let's consider
 1-dimensional analogue of notorious  two slits experiment (TSE)
   of  QM   \cite {Schiff,Fey}.
We shall regard  
% the initial state  $|g_0\}$, which support  $E_x$ in some RF
%       which is the sum of two substates $g_{1,2}$ 
the system  of $n_s=2$ point-like $m$ sources (bins) with
$Dx_{1,2}$ width cited in $x_{1,2}$.
% for which the state $|g(t)\}$ is the studied signal. 
%   the initial $m$ density $w^0_s$
%  is decomposed as $w^0_s(x)=\sum w^0_{i}(x)$,
%   where $w^0_i(x)$ is located in $Dx_i$.
% After  $g_0$ preparation at $t=0$ presumably 
%  $m$ doesn't interact with any other
% object and evolves freely.
%       $g_0$ can be regarded  as the source $S(g)$.
% the resulting density $w(x,t)=\Xi^f_t g_0$.
% The fuzzy map $\Xi^f_t$ in principle can
%    project   the internal fuzzy structure of the source
% $S(g_0)$  to the distribution of signal density $w(x,t)$,
%                     and due to it SS effect  will appear. 
%
Consider first the probablistic mixture $g^{mix}_0$
of  $g^0_{1,2}$ states, localized in $Dx_{1,2}$ respectively,
 in that case, the weight $w^0_i=P_i$ where $P_i$ is the probability
for $m$ to be in $Dx_i$, the density of $m$ sources is: 
\begin {equation}
  w^0(x)=\sum \limits^{n_s} w^0_i\delta (x-x_i)  \label {D1} 
\end {equation}
over $m$ ensemble  (we regard the mixtures in which $w^0_{1,2}$
are  the same or don't differ much).  In each
individual event $m$  is  emitted definitely  by $Dx_1$ or $Dx_2$
 at $t_0$, therefore  $g^{mix}_0$ algebraic  structure
is described by  the following proposition:
$$
LP^{mix} := \quad m\in Dx_1 \, .or.\, m \in Dx_2
$$
%   each $w_i$ corresponds strictly to one of this outcomes, i.e.
%   is random source $S^m_R$.
Consequently, the resulting $m$ distribution over this ensemble at any $t>t_0$
 will be the additive sum: 
$$
 w_{mix}(x,t)=w_1(x,t)+w_2(x,t)
%     P_1 w^r_1(x,t)+P_2 w^r_2(x,t)
=\sum w^0_i \Gamma_w(x-x_i,t)
$$
For SS illustration the most interesting is the case
when $w_{1,2} (x,t)$ intersect largely,
i.e. for  $L_x=|x_1-x_2|$ it should be 
$L_x \le \sigma_x(t)$   where $\sigma_x(t)$ is $w_{1,2}$
dispersion. For our toy-model
% when $w_i=w^D(x-x_i,t)$
 it holds if $L_x \le \kappa t^{\frac{1}{2}}$.
 The rate of $w_1,w_2$ overlap can be estimated as: 
$$
            R_w=2\int \sqrt {w_1 w_2} dx
$$
%   where $w^r_{1,2}$ are the densities from the single event.
%   where each $w_{1,2}$ corresponds to 
%   emission  by one of local sources $Dx_{1,2}$. 
and it shouldn't be much less than $1$.

Now consider the pure fuzzy state $|g_0\}$ for which $m$ coexists
simultaneously in both bins $Dx_i$ with the same weights $w^0_i$,
more precisely, $g_0$ is supposed to be the superposition of $g^0_i$ states
 of regarded mixed ensemble ( exact FM definition of state's superposition
 will be given below).
 For this pure  $m$ state $|g_0\}$ the following proposition
 describes $m$ source structure:
$$
 LP^s  := 
%   m \in (Dx_1\cup Dx_2) \, .and.\, 
m \wr D x_1 \, .and. \, m \wr D x_2
$$
where $m \wr Dx_i$ means that $m \wr x_a; \, \forall x_a \in Dx_i$.
% But as follows from $LP_f$ in case of
In this case,  $LP^{mix}$ and $LP^s$ are incompatible:
 $$
LP^{mix}.and.LP^s = \emptyset
$$
%for an arbitrary  proposition $LP^e$ which describes also some
%$m$ signal. 
 The incompatibility of $LP^s,LP^{mix}$ indicates that
the signal of fuzzy source $S$ can't be decomposed into the sum of signals
from  local  sources $Dx_{1,2}$.
For such source's system
 from $w(x)=w_{mix}(x)$ follows $LP^{mix}=.true.$
 with definiteness. Hence, if the  resulting distribution $w_s$
  to decompose as:
$$
                 w_s(x)=w_p(x)+k_w w_{mix}(x)
$$
where $w_p \ge 0$ is arbitrary, it follows that $k_w$=0, i.e. any $w_{mix}$
 content in $w_s$ is excluded.
% In  minimal FM $w_s$ is local in a sense that
% so that the resulting distribution $w_s(x,t)$ is of approximately the same
%  width as $w^{mix}$.
In other words,   $w_s(x,t)$  should have such form that it  makes
 in principle impossible 
to represent $w_s$ as the sum of two components, each of them
 describing $m$ signals from $Dx_{1,2}$  sources.
If $w_s, w_{mix}$ supports in $X$ mainly coincide, such $k_w$ value is possible
only if  $w_s$ oscillates around $w_{mix}$ and
in  one or more points $x_j$ where $w_{mix}(x_j) \ne 0$
 it gives $w_s(x_j)=0$. Plainly, such picture describes
 the interference patterns similar to observed for QM superposition. 
%
% It should be maximally
% different  from the mixture $w_{mix}$, so its content in $w_s$ 
% should be minimal
$w_s$ can be decomposed as:
$$
       w_s(x)=w_1(x)+w_2(x)+w_n(x)
$$
where $w_n$ is  the nonlinear term.
For our toy-model it can be equal:
$$
   w_n(x) \sim 2\cos[r_D(x -\frac{x_1+x_2}{2})]
 [w_1(x) w_2(x)]^{\frac{1}{2}}
$$
where $r_D$ is arbitrary but $r_D \gg \kappa \sqrt{t}$      \cite {Vax}.      
%  $\sqrt{w^0_1w^0_2}$.
%    Such $w_s$ joint
%  dependence on $w^0_{1,2}$ is the characteristic
%  feature of the fuzzy map. 
%
One should define also SS measure i.e.
 the criteria of signals separation - $R_{ss}$ for the
evaluation of smearing rate; depending on it
$R_{ss}$ can vary  from $0$ for $g^{mix}$  to $1$ for fuzzy state
with maximal SS. 
% In principle, an alternative measures can be used, but for our problem
% they lead to effectively same results. It can be shown that
% in any realistic situation $R_{ss} \le R_w$, so to get the
% maximal SS  it's necessary  that $R_w \rightarrow 1$ also.
%   If $C_x \rightarrow 0$
%   TSE mixed and fuzzy distributions simply coincide.
The general $R_{ss}$ ansatz is quite complicated  \cite {Vax},  
but $R_{ss}$ will be used in our formalism  
only in the  asymptotic limits
when $R_{ss} \to 0$ or $1$.
%  in that case one can take $R_{ss} \simeq R_w$.
In FM framework, $m$ free evolution SS effect should respond
to maximal $R_{ss}$ value, because in FM no information about $m$
path  from the source $|g_0\}$ exists at all. Otherwise, it would mean
that the additional information about $m$ source (path) is produced
 stochastically during $m$ evolution,
 but it's impossible, because of FM reversibility. 
   
% For example, if   $w_i= w^D(x,t)$ of
% classical diffusion (but with resulting $w_n \ne 0$),
%   $w_i \sim w^0_i\exp{-\frac{(x-x_i)^2}{\sigma^2_x(t)}}$, 
% then this property is obvious.

The similar  SS effect should be expected for 
complete FM formalism, hence it's
instructive to exploit whether Fuzzy Geometry prompts some indications for
SS geometric  scale  characterized by  $\sigma_x (t)$.  By itself,
 Fuzzy Geometry
doesn't contain any length parameters which can be put in correspondence 
to $\sigma_x$.
Really,  the  fuzzy point $b_j$ described by $w_j(x)$
possess the obvious scaling properties for $w_j$ dispersion.
 From that it's quite natural
to expect that in 1-dimensional FM the influence of source $g_0^i$
 on the state $|g(t)\}$ in point $x$ is independent on $|x-x_i|$.
Hence minimal FM also should show the scaling behaviour, which permit 
to omit any length parameters settling
 $\sigma_x(t) \to \infty; \, \forall t$.  
In relativistic theory,  the
 dispersion $\sigma_x (t)$ is restricted
by the maximal velocity $c$, so that $\sigma_x \le ct$.
In   nonrelativistic case, 
 nothing forbids to choose  FM ansatz for point-like source $x_i$ 
% with $\sigma_x(t) \rightarrow \infty$ at finite $t$, and 
such that  at $x \rightarrow \pm \infty $,
 $\lim w(x-x_i,t) \ne 0$ (or the limits don't exist) \cite {Schw}.
 This $w (x-x_i,t)$ property is called   $x$-limit condition,
in our toy-model it fulfilled only for $t \to \infty$.
 Then
$w (x)$ should be Schwartz  distribution (generalized function) \cite {Vlad}.
% and the  class of distributions which obeys it is denoted $W^x$.
% In this case, for system $n_s=2$ $R_{ss}$
% can be independent of $L_x$, even for
% $L_x \rightarrow \infty $,  so such theory doesn't need the additional
% length  parameters.
 Such $|g\}$ evolution at first seems quite exotic,
 remind yet that in QM the point-like initial state in 1-dimension
evolves analogously \cite {Fey}.

Consider now system of $n_s=2$ sources with  particular $x_{1,2},w^0_{1,2}$,
each state  $g^0_{1,2}$ evolves into $w_{1,2}(x,t)$
 which satisfies to $x$-limit condition.  If $|g_0\}$ is their superposition 
 then the resulting $w_s(x,t)$ should also satisfy to it.
  $\bar{x}(t)$ and higher $x$-moments are undefined
 for such $w_s(x)$ and 
in that case only $w_s(x)$ form can depend on  
 FM dynamics. In FM this  
 $w_s(x)$ should  respond also to the maximal SS, i.e. $R_{ss} \to 1$.
Then   $w'_s(x,t)=w_s(x+a_x,t)$ also responds to it  for an arbitrary $a_x$,
because $R_{ss}$  depends of $w_s$ form only.
If $R_{ss}$ maximality is the only condition of $|g(t)\}$ consistency,
then $w'_s$  also can be the solution for some $g_0$ state which
is the superposition of $g^0_i$.
% If $\sigma_x(t)$ is finite and $\bar{x}(t)$ is well defined,
%  then it unambiguously stipulated by the initial state and its dynamics;
% the example is the classical diffusion. 
%components, but the alternative solutions with an
% arbitrary $a_x$ are consistent also.
This conclusion is especially obvious if $w_i(x,t)$ are
practically  independent of $x$, in our toy-model it occurs for
$t \to \infty$.
% $w_s$ form can be characterized numerically by its
% fourier-transform and other methods which details are unimportant here
% \cite {Ed}.
%   At first sight this assumption
%  seems preposterous but the calculations results presented below  
%  supports it. 
 Hence the resulting $a_x$ value should be defined by the initial
$|g_0\}$ state; this considerations evidence that
 beside $w(x),\, |g\}$   includes
  at least one more degree of freedom.
Since $a_x$ depends on $|g\}$ both in $x_1$ and $x_2$,
 it is sensible to assume
that it can be represented as the  correlation field $g^2_i(x_1,x_2)$
introduced above.
%  Really, as was noticed above, for  $n_s=1$ $g_0$ is equivalent to
%  the ordered point $x_0$, so in minimal FM it's sensible to assume
%  that such  $g_0$ doesn't have any internal structure
%  which can be reflected to $\bar{g}_a$.
%   Indeed, obtained $w(x,t)=const$ for $n_s=1$
%  demonstrates that such $g_0$ can't have any free $\bar{g}^a$
%   parameters  on which $w$ can depend.
%  In this case  
% the meaningful $\bar{g}^a_0$ has a sense only for $w_0 \neq  \delta (x-x_0)$,
% i.e.  expressed as the binary correlation $K^f(x,x')$ (or
% higher order correlation).
 In  minimal FM for arbitrary state $|g\}$  it can be an  arbitrary real
 function of two variables $g^2_1=K^f(x,x')$ which is  continuous or has
the finite number of breaking points.
Consequently,  for $n_s=2$ system  $a_x$ is some  function:
 $a_x=f^f[K^f(x_1,x_2)]$.
%  where $F_k$ is some $K^f$ functional.
%   For $K^f$ - real scalar 
 If to choose the gauge: $\forall x_b;\,K^f(x_b,x_b)=0$, then 
 regarding the  fixed $x_c$ as the parameter we obtain:
$$
  K^f(x_d,x_c) =\int\limits^{x_d}_{x_c}
 \frac{\partial K^f(\xi,x_c)}{\partial{\xi}} d\xi
$$
where $\frac{\partial K^f(\xi,x_c)}{\partial{\xi}}$ is integrable function
and has the finite number of breaking points.
From that it follows:
$$
K^f(x_d,x_e)=K^f(x_d,x_c)-K^f(x_e,x_c)
$$
 Therefore  $K^f$ is, in fact,  the function of one observable:
$\lambda(x)=K^f(x,x_c)$. Because of it
 $g$ can be regarded as the local field
$E^g(x)=\{w(x),\lambda(x)\}$.     
%  For FM with  maximal SS $w$ evolution is quite singular for $n_s=1$
%   Note, that $x$-limit condition closely related to $K(x)$ appearance
%  but isn't exactly the same thing. 
% To simplify the evolution ansatz,
% we exploit   the map $O E^g(x) \rightarrow g(x)$
% which parameters are defined below. It transforms $E^g$ into the 
%  expressed as: $\{ w(x),K(x) \} \rightarrow \{
One can to change it for the symmetric $|g\}$ representation by the
complex  function $ g(x)=w^{\frac{1}{2}}(x) \exp{ic_{\lambda}\lambda(x)}$ where
$c_{\lambda}$ parameter will be calculated below.
% g_1(x)+ig_2(x)$, where
%  $g_{1,2}$ are the real function
% $m$ density - $w(x)=F_w(g(x))$, where $F_w$ is  an  arbitrary
% analytic  function.
%  The relation of this $g$ representation with Superposition Principle is
%  discussed below.
%If $g$ is the local field, it's natural  to assume that 
In this case, $w/g$ zero-equivalence holds:
$w(x)=0 \,\Leftrightarrow \, g(x)=0$ and the same is true for $w,g$ limits
at $x \rightarrow \infty$.
%     which will be important in our calculations.
%  $w/g$ zero-equivalence and
In this case, if  $x$-limit condition fulfilled for $w(x,t)$, then it's true
also for $g(x,t)$ which should be also Schwartz distribution.
 %Of course, the alternative reasons for the additional $g$
% degrees of freedom -  $K(x)$
% appearance can exist, but in FM, as will be shown below,  
% the proposed explanation is    the most appropriate one.
% Note that $K(x)$ can be applied straightforwardly only for
% for the pure states with $l^g \ne 0$, for the mixtures
% its use  is more complicated.
 Note that  for the finite $\sigma_x(t)$ such  dependence between
$a_x$ and $m$ state $|g_0\}$ can exist in our theory but the corresponding
 ansatz  will be more complicated.
In general, one should be careful with the interpretation of $w(x,t)$
distributions as the measurable distributions of physical parameters,
 yet in the discussion of QM foundations  it's admissible
to regard them  as
the standard, normalized functions,
as was demonstrated in \cite {Fey}.
Below  this problem will be reconsidered in detail.

\section {Particle's Evolution in Fuzzy Dynamics}

% After the semiqualitative FM toy-model consideration
% we can turn to the fuzzy states $|g\}$  
%  evolution formalism.
 From the previous discussion  we concede that in FM 
the state of particle $m$ in 1-dimensional space $X$
is described by normalized complex function $g(x,t)$;
 for free $m$ evolution from point-like source
it satisfies to $x$-limit condition.
% for which $w/g$ zero-equivalence holds. 
% In  FM framework,  free evolution
% is assumed to be reversible and invariant under $X,T$ shifts.
%  it  will be shown  that
% this  assumptions are enough 
% to calculate $m$ free evolution consistently.
%  it differs from the  standard QM approach
%  in which more axioms  are necessary.
% After the semiqualitative discussion we come to FM evolution calculations;
%  Our main  result for FM free evolution can be formulated as follows:\\
% { \bf Theorem:}  
% We shall assume also that
% for $m$ free reversible evolution  $g(x,t)$ and its
% Fourier-transform $\varphi(p,t)$ are continuous  for  
%   $t \ge t_0$.
%   and $w/g$-zero equivalence holds for $g$, then
% We'll argue that under this conditions  
% $g$ would  evolve in accordance with
% Schr$\ddot{o}$dinger free evolution operator
% ${U}_s(t)$ and $F_w(x,t)=|g(x,t)|^2 $. 
% It would permit also to find the 
% undefined $O$  map parameters.
In general,   
 $g(x,t)$ 
 reversible evolution is described by the parameter-dependent 
unitary operator
$\hat{U}(t)$, so that: $g(t)=\hat{U}(t)g_0$.
% $\hat U$ corresponds to $\Xi^f_t$ in $g(x)$ representation of $|g\}$.
It possesses the  properties of group element:
$$ 
 U(t_1+t_2)=U(t_1) U(t_2); \quad  \forall t_{1,2}
$$
 therefore  $m$ free evolution can be expressed as
$\hat{U}(t)=e^{-i \hat{H}_0 t} $ where $\hat{H}_0$ is an arbitrary constant  
 operator \cite {Ber}. It isn't supposed beforehand to be linear,
but we start from the consideration of linear $H_0$,
 the obtained results will help us to analyze the nonlinear case.
%
% Meanwhile,  the space shift operator $V$ is equal to:
% for  the space shift of $g(x,t)$.
% from that $V(a)=e^{iap}$ when acting on
% $\varphi(p,t)$ which is $g(x,t)$ Fourier-transform
% \cite {Schiff}.
 The free $g$ evolution is invariant 
relative to  $X$ shifts performed by the operator
  $\hat{V}(a)=\exp({a\frac{\partial}{ \partial x}})$.
 Because of it, $\hat{U}(t)$ should commute with 
$\hat{V}(a)$ for the arbitrary $a$. It's equivalent to the
relation $[\hat{H}_0,\frac{\partial}{ \partial x} ]=0$, 
from which follows that $\hat{H}_0$ in $p$ representation 
% for $\varphi(p,t)$
 is an arbitrary function of $p$: $H_0=F_0(p)$.

% $g$ fourier-transform $\varphi(p,t_0)$

% with $\varphi$ of (\ref {DD11})   $H_0=F(p)$ and $ a=1;\, \beta=0$.
% if  $g(x,t)$ is  continuous or  at least
%  its Fourier transform $\varphi(p,t)$
% is continuous  for any $t \ge t_0$.
 Consider now  the initial point-like state $|g_0\}$ inducing $m$ density
 $w^0= \delta (x-x_0)$; 
%  the $w/g$ zero-equivalence at $t=t_0$ 
% the obvious condition for 
 we put in correspondence to it the unnormalized function 
 $g_0(x)=\exp ({i\alpha_0})\delta(x-x_0)$
 where $\alpha_0$ is an arbitrary real
number.
The proper $g_0$ normalization will be regarded below,  at this stage
it will introduce the unnecessary complications
 but wouldn't change $g$ ansatz principally.
Then from  $\delta(x-x_0)$ fourier transform
 $\varphi_{\delta}(p)=\exp ({ipx_0})$ it follows that $g$ fourier transform is
equal: 
 $$
       \varphi(p,t)= U(t) e^{i\alpha_0}\varphi_{\delta}=
  e^{-iF_0(p)(t-t_0)+ipx_0+i\alpha_0}
$$
%      From that for $n_s=1$
%   it follows $g(x-x_0,t)=g(x_0-x,t)$ for $n_s=1$.
%  We'll show that  the  theorem premises define $g$ distribution
%  evolution unambiguously.
 below  $x_0=0, t_0=0$ settled.
The transition
 $ \delta(x)\rightarrow g(x,t)$  develops  continuously without breaking
 points  if $g(x,t_j)$  constitutes $\delta$-sequence, 
i.e.  $g(x,t_j)\rightarrow \delta(x)$ for any
sequence   $\{t_j\} \rightarrow +0$ \cite {Ed}.
 It means that for an arbitrary function $\chi(x)$,
 which belongs to the class of main functions  \cite {Vlad},
 one has: 
$$
   I (\chi,t)=\int\limits^{\infty}_{-\infty}
 \chi(x) g(x,t) dx \rightarrow \chi(0)
$$
at $t \rightarrow +0$.
 This condition is fulfilled only if $g(x,t)$ has $t=0$ pole,
so that $g(x,t)$ can be decomposed as: $g=g_s g_a$ where
 for the substitution
$z=\frac{x}{f(t)}$ one obtains:
%  $g_s$ class described above,
$$
    g_s(z,t)=\frac{1} {f(t)} e^{i\gamma(z)} ,
$$
with an arbitrary, complex $\gamma$;
 $f(t)\rightarrow 0$
at $t \rightarrow +0$.  
% at $t \rightarrow +0$ \cite {Vlad}.
% which means that $g$ contains $f(t)$ pole \cite {Ber}. 
 $g_a$ is an arbitrary, nonsingular function with
 $g_a(x,t) \rightarrow  1$
  at $t \rightarrow +0$, so it can be neglected in this limit.
If in this limit:
$$
\int\limits^{\infty}_{-\infty}
  g(z,t)f(t) dz \rightarrow 1
$$
 then under this conditions
$g(x,t) \rightarrow \delta(x)$ at $t \rightarrow +0$.
%
% Consider the function $g_s$ of the form:
% where $z=\frac{x}{f(t)}$ with the  complex $f, \gamma$.
%  Let's suppose also that at $t\rightarrow +0$, $f(t)$
%  approximated  as $f(t)= d_rt^r+o(t)$ with $r>0$
%  and $d_r$ - an arbitrary constant.
%   As will be seen below  $f(t)$ which doesn't
%  approximated in this way, alike $f=t \ln t$ are unsuitable.
%
%  \gamma=\gamma_z(z)+O_{\gamma}(x,t) =
%  [\frac{x^2}{f(t)}  ]^b+O_{\gamma}(x,t)=z^b+O_{\gamma}  $ ? (r) (r)"е- ¤
%
After $z$ substitution $g$ Fourier transform $\varphi$
alternatively can be  represented as:
\begin {eqnarray}  
\varphi'(p,t)= c_0 \int \limits^{\infty}_{-\infty} dz
 e^{i\gamma(z)+izpf(t)}= \exp^{-i\Lambda[pf(t)]}
              \nonumber \\
\end {eqnarray}
% Decomposing $g_a$ as the row in $t$, 
 From the equivalence  
  $\varphi (p,t)$ and $\varphi'(p,t)$
one obtains  the equation:
\begin {eqnarray}
 \varphi(p,t)=e^{-i F_0(p) t} = \exp^{-i\Lambda[pf(t)]+\alpha_e}
%  =c_0[1 -i F_0(p) t]=\varphi(p,t)
  \label {DE2}
\end {eqnarray}
from which follows $F_0(p)=\frac {p^s}{2m_0}, f(t)=d_rt^r$, $\alpha_e=0$,
with $rs=1$, where $m_0,d_r$ are
an arbitrary  parameters.
%  From that one finds $g_a(x,t)=1$, 
%  $\Gamma(pf,t)=\Gamma(pf,0)$, and  the former equation
% holds true at any $t$.
 If  $H_0=F_0(p)$ is regarded as
   $m$ free Hamiltonian,
then from its  symmetry properties and the energy positivity
it follows that $m_0 >0$ and
 the consistent $s$ values are only  the natural even numbers.
%  and so $r \le \frac{1}{2}$.
%\cite {Schiff}.

% The detailed proof can be given
% only after the FM Hamilton formalism will be formulated 
% in the final part of our paper. Sketching it beforehand, 
% only in  case $s=2$  linearity of $m$ velocity transformations:
% $\dot{x}'=\dot{x}+v$ necessary for RF transformation
%  is  compatible with the solution of Heisenberg equation: $\dot{x}=i[x,H_0]$
%

Let's consider first the case $s=2$,
 it follows that the free Hamiltonian is  $H_0=\frac{ p^2}{2m_0}$;     
%  $U(t)$  is the unitary operator
%for  the real $m_0$.
%
%  It gives also $  f(t)=\sqrt{c_m t}; \\
%  Eventually, $c_0= e^{i\alpha_0}\sqrt{\frac{-i2\pi} {m_0}}$;
%
%  up to the arbitrary, real $\alpha_0$ ??????????????
%  If  to demand that the shift in $p$-space
%  $p'=p+f_m(m_0)v$ for an arbitrary, constant  $f_m$
%   and  the parameter $v$  corresponds to 
% Galilean coordinate transformations: $t'=t,\,x'=x+vt$,
% where $x_G$ is the space shift between two RFs, and 
%  $H_0$ ansatz is invariant under this transformation,  then this conditions
%   are satisfied only for $s=2$.
%
 for point-like state $g_0(x) = e^{i\alpha_0} \delta(x-x_0)$ one
obtains :
$$
   \varphi(p,t)=e^{-\frac{ip^2 t}{2m_0}}
$$
which  in $x$-representation results in:
\begin {equation}
    g(x,t)= G(x-x_0,t)  e^{i\alpha_0} = \sqrt{\frac{m_0}{-i2\pi t}} 
 e^{\frac{im_0(x-x_0)^2} {2t}+i\alpha_0 }
             \label {DD23}
\end {equation}  
hence for  positive $m_0$ value 
 $G$ coincides with QM  free  propagator for particle with mass $m_0$
up to arbitrary constant \cite {Fey}.
% it defines $\gamma,f$ completely.
%  Such $m_0$ value leads to $g(x,t) \ne 0$ at $x \rightarrow  \pm \infty$
%  and satisfies to $x$-limit condition.
% Then in this formalism   $m_0$ can be interpreted as the  particle $m$ mass ;
% note that for an imaginary $m_0$ such ansatz describes the classical diffusion.
%  we'll regard it for the arbitrary $g_0$.
%
%  Only the real $f,\gamma$ results in  distributions $g(x,t) \ne 0$ at
%  $x \rightarrow \pm \infty$; for  the imaginary $\gamma$
%   $g$ evolution is the classical diffusion analog \cite {Ber}.
% $g_0$ for the arbitrary $n_s$ can be written as :
% $$
%  g_0=\sum^{n_s} \sqrt{w_l^0}\delta(x-x_l)e^{i\alpha^0_l} 
% $$
% where $\alpha^0_l=K(x_l)$ are  an arbitrary real constants.
% From $U(t)$ unitarity  evolution defined by $G(x-x_l,t)$ of (\ref {DD23}).  
% Thus an $n_s$ initial state  besides $N_l$  depends only on the $n_s-1$
% parameters $\alpha_l^0-\alpha_i^0$ which defines the correlations
% between the fuzzy sources.
% Obviously,  one can transfer from the sum with
%  $n_s\rightarrow \infty$ to  
If to accept that $G$ is FM free $m$ propagator, then
  an arbitrary normalized function describing the initial state
 $g_0(x)=\sqrt{w_0(x)} e^{i\theta(x)}$ will evolve as: 
\begin {equation}
 g(x',t)= \int G(x'-x,t) g_0(x)dx
=\sqrt{ \frac{m_0}{-i2 \pi t}}
\int e^{\frac{im_0(x'-x)^2}{2t}} g_0(x)dx    \label {DD33}
\end {equation}
which coincides with the free $g_0$ evolution in QM 
formalism \cite {Fey}.
%   and $G$ is free Feynman propagator \cite {Fey}.
% For that  $U(t)$ evolution,
%
For such evolution ansatz one finds that the integral form
$\int |g(x,t)|^2dx$ is time independent and equal to $1$;
in this case, $w=|g|^2$ satisfies  to $m$ flow conservation
equation.
% Therefore $w(x)=F_w$
% can be chosen as $m$ universal density; in particular, it
%permit to chose $c_0=1$.
%  which induces
%  also $\langle g_1|g_2\rangle$ scalar product choice \cite {Schiff}.
Note that for $s=2, \, g(x,t) \neq 0$ at $x \rightarrow \pm \infty$,
i. e. satisfies to  $x$-limit condition, as  minimal FM ansatz assumes.
Yet it violated for free Hamiltonian with $s \geq 4$,
 in this case, $g(x,t)$ asymptotic
can be  calculated \cite {Fed} at   $ x \to  \pm \infty$:
$$
   g(x,t) \simeq\frac{c_g}{ t^{\frac{1}{s}} }
(\frac{ t^{\frac{1}{s}} }{x} )^{\frac{s-2}{2(s-1)}}
\exp^{i\frac{s-1}{s} m^{\frac{1}{2(s-1)}}
 t^{-\frac{1}{2(s-1)}} {x}^\frac{s}{2(s-1)}}
$$
with $c_g$ - arbitrary constant.
In particular, for $s=4\,$, $|g| \sim \frac{1}{|x|^\frac{1}{3}}$.  
Therefore $g \to 0$ at $x \pm \infty$,   so it contradicts to $x$-limit
condition, because of it the important  assumption of minimal FM is violated
for $s \ge 4$.
% In addition it can be shown that for $s \geq 4$
% it's impossible to construct $w(x)=F_w(g)$  as  a nonnegative,
% local $g$ form   which obeys the flow conservation. 
% Therefore, $x$-limit condition is quite important,  because it selects
% the unique ansatz for free FM evolution operator.

Let's consider now the general case of FM free evolution, which
doesn't demand, in principle, that $\hat{U}(t)$ operator should be linear.
However,  $m$ evolution is supposed to be reversible, so $\hat {U}(t)$
must be unitary. The thorough investigations of nonlinear
Schr$\ddot o$dinger-type operators have shown that such physically nontrivial
 operators are
nonunitary \cite {Gold}. In accordance with it, we shall demonstrate  
that unitary free $m$ evolution can't be induced by nonlinear  Hamiltonian
$\hat{H}_0$. Here we only sketch the proof leaving some mathematical
details for the future study. Consider an arbitrary $m$ normalized
 state $|g\}$  in $p$-representation:
$$
   \varphi (p,t) = [w_p(p,t)]^{\frac{1}{2}} e^{i\beta(p,t)}
$$
where $\beta$ is real function, $\varphi$ obeys the equation:
$$
 -i\frac{\partial \varphi}{\partial t}=\hat{H}_0 \varphi
$$ 
Since free $H_0$ is invariant relative to the space shifts, then   
$\langle p^n \rangle$ are constant $\forall n$, from that it follows that 
$\frac {\partial w_p}{\partial t}=0$. It transforms the latter equation into:
\begin {equation}
                -i \frac{\partial \beta}{\partial t}\varphi
              =\hat{H}_0 \varphi     \label {XWX}
\end {equation}
i.e. $H_0$ action results in multiplication of $\varphi$ on
some function $F(\varphi, p)$.  $H_0$ is the constant operator,
hence $F$ can't depend on $t$ as parameter but only via $\varphi$. 
The obvious solution is $\hat{H}_0=F(p)$,
and  this  just corresponds to the regarded linear ansatz with $F=F_0(p)$.
%       F_p(p) = \frac{\partial \beta}{\partial t}
% from which immidiately follows that $H_0$ is the linear operator.
The simple analysis shows that no other
consistent and physically interesting  $\hat{H}_0$ solutions exist.
Such solutions or violate $T$-shift invariance of FM, or
don't permit to define $m$ energy ansatz unambiguously. 

Now the normalization of $g(x,t)$ states from point-like $m$ sources
will be considered, it's valid also for QM formalism where
this aspect often missed \cite {Fey}. This problem is quite trivial,
so in place of universal derivation we shall regard it for  particular
$\delta$-sequence. Plainly, the state of point-like source $g_0(x)$
 should be the 
limit  of physical normalized states of very small width.
 Namely, it can be the sequence of initial states:
$$
\eta_{\sigma}=
\frac{e^{-\frac{x^2}{2\sigma_x^2}}}{{\pi}^{\frac{1}{4}}\sigma^{\frac{1}{2}}}
$$
for $\sigma \to 0$; the resulting function
 $\delta^{\eta}(x)=\lim \eta_{\sigma}(x)$ called  the squire root of
 $\delta(x)$.  $\eta_{\sigma}$
%  its norm $\int|\eta_{\sigma}|^2dx$  is 1 and its width tend to $0$. Its
 density  $w_{\sigma}(x)$
%=|\eta(x)|^2
%=\frac{e^{-\frac{x^2}{\sigma_x^2}}}{{\pi}\frac{1}{2}{\sigma}}  
%% $$
has the norm $1$ and the limit  $\delta(x)$,
 as  expected for the state
of point-like source. Hence it seems consistent to choose
 $g_0(x)=e^{i\alpha_0} \delta^{\eta}(x)$ as the pointlike state in FM
(and QM also).    
% The  evolution equation for its fourier-transform  
% coincides with  \ref {DE1} and its solution can be
Its fourier transfom expressed as:
$$
     \varphi_{\eta}(p,t)=\lim_{\sigma \to 0}
 \frac {\sigma^{\frac{1}{2}}}{(2\pi)^{\frac{1}{4}}}
 e^{i\alpha_0-i\frac{p^2t}{2m_0} -2\sigma^2 p^2}
$$
 also has the norm 1 at any $t$. $\varphi$ 
 describes the normalized constant distribution of
$m$ density on  $p$ axe. If to substitute such $g_0$ into (\ref {DD33}),
the resulting $g(x',t)$ will have norm $1$ at any $t$. It stresses that
the propagator $G$ isn't the physical state of particle $m$.
However, all FM results obtained above don't depend on this renormalization
and stay unchanged, becouse such renormalization is, in fact,
 the multiplication of $g_0(x)$ and $g(x,t)$ on the infinitesimal constant.
% In general, the physical 
% observables, corresponding to numerical values in QM and FM, 
% respond to the states bilinear form  of the kind
% $ \psi^* \hat{A} \psi $. The states by itself can't      
% induce the real numbers in this way, so they shouldn't be 
% the generalized functions with neccesaty.
% and  it seems to explain
% the obtained ansatz for localized states $g_0$.

% One can consider them  also as the limit of
%the normalized observables  distributions; for example, $w(x)=const$
%is the limit of Gaussian with $\sigma_x \rightarrow \infty$.

  It turns out that  the obtained $\hat{U} (t)$
 ansatz coincides with QM Schr$\ddot o$dinger evolution
operator for  free particle evolution.
% Moreover, it agrees with the simple picture proposed in our FM toy-model.
 The analogous results for QM
are obtained  in the theory of  irreducible  representations,
but  in that case they are based on more complicated axiomatics,
which, in particular, includes the axiom of 
 Galilean invariance \cite {Schiff}.
In distinction, FM doesn't assume  
 Galilean Invariance of $g$ states in different RFs but only the invarinace
relative to the space and time shifts.
 It acknowledged in Quantum Physics that the classical
massive objects, including physical RFs, can be regarded
as the quantum objects in the limit  $m_0 \to \infty$ \cite {Schiff}.
If such approach is correct in FM framework also,
then regarding $m$ with $m_0 \to \infty$ as RF, 
   Galilean transformations 
  can be derived from the obtained FM ansatz for $H_0$.
Of course, this hypothesis needs further investigation,
 but in this approach it seems consistent.

Now  Hamilton formalism for FM can be formulated consistently. In our theory
% from $X$ space shift symmetry it follows that
  $m$  momentum is the  operator
 $\hat{p}=-i\frac{\partial}{\partial x}$ \cite {Schiff}
in $x$-representation
%   with $[\hat{p},x]=i$,
%   so that  $\bar{p}=\frac{\bar{\partial\alpha}}{\partial x}$ \cite {Schiff}.
%  Yet we know that  the linear complex functions evolution described by
%   Schroedinger equation (SE) of QM, where Hamiltonian $H$ becomes Hermitian
%  operator.  It guarantees
%  also $m$ flow conservation and restores classical limit for
% arbitrary $H$ \cite {Schiff}.
and  the free
 Hamiltonian  $\hat{H_0}=\frac{\hat{p}^2}{2m_0}$. 
 In FM the natural $U(t)$  generalization
 for the $m$ potential interactions $V_m(x)$ is:
 $\hat{H}=\hat{H}_0+V_m(x)$. From obtained relations it results in
Schr$\ddot{o}$dinger equation for $g$;
the general path integral ansatz for $g$ can be obtained
 by means of Lagrangian $\cal L$  derived from $\hat H$
 for the given  $V_m(x)$ \cite {Fey}.
%  The correlation tensor of minimal FM model corresponds to
%  a quantum  phases differences $K(x,x')=\alpha(x)-\alpha(x')$
% The quantum phase $\alpha(x)$ properties acquires the natural description
% in FM framework:
% the real physical parameter is $K(x) \sim K^f(x,x')$ -
% the fuzzy correlation between 
% $x,x'$, and $\alpha=K(x)$ is its local  $x$
%  representation  which is ambiguous up to $2n\pi$.
Any normalized function $g(x)$ admits the orthogonal decomposition
on $|x_a\rangle=\delta(x-x_a)$, and $|x_a\rangle$ set constitute
the complete system \cite {Ber}. Therefore 
$|g\}$ set $M_s$   is equivalent to complex 
rigged Hilbert space $\cal{H}$  with  the scalar product
 $ g_1* g_2 = \int g_1^* g_2 dx$.
Consequently, our theory doesn't need Superposition Principle as
the independent axiom, it follows from already exploited FM axioms.
In FM  $x$   is $m$ observable and it's sensible
to suppose that $\hat p$ and any Hermitian operator function $\hat {Q}(x,p)$
 is also $m$ observable. For any such $Q$ there is
the corresponding complete system of orthogonal eigenvectors $|q_a\rangle$
 in $\cal H$.
It permit to assume that for FM measurements of observables 
 QM Reduction (Projection) Postulate for an arbitrary observable $Q$ 
can be incorporated in FM copiously \cite {Schiff}.
% so that  the 
% particular outcome of $Q$ measurement $q_i$ 
% appears with the probability $P=|\langle g| q_i\rangle|^2$  \cite {Schiff}.
The important FM advantage is the relative independence of its formulation from
realm of Classical Physics. Copenhagen QM interpretation claims that
QM can be formulated consistently only on  preliminarily postulated
definitions of Classical Physics, it seems that FM formalism is, at least,
essentially less connected with them. 
%
% For the conclusion, it was shown that  1- dimensional $M^F_{s-t}$ Fuzzy
%  Geometry
% induces $m$ massive particles dynamics which under simple assumptions
% coincides with QM Schr$\ddot{o}$dinger dynamics.
%
  Generalization of
 FM formalism  on 3 dimensions is straightforward,
% and will be regarded in the
% forcoming paper; here we  sketch only the main points.
%We assume that the signal $g(\vec{r},t)$ from the point-like source 
% $w_0=\delta(\vec{r}-\vec{r}_0)$ possess the spherical symmetry.
the only novelty is that $|g\}$
 correlation  between two points $K^f({x}_{1,2})$, defined
 in  previous chapter, now acquires the form:
$$
  K^f(\vec{r}_1,\vec{r}_2) =\int\limits^{}_{l}
 \frac{\partial K^f(\vec{r},\vec{r}_2)}{\partial{\vec{r}}} d\vec{l}
$$
and supposed to be independent of the path $l$ over which it calculated.
In this case, $|g\}$ quantum phase $\alpha(\vec{r})$ is defined
unambiguously.  
% Analogously to 1-dimensional case $n_s=2$ we assume that for
% the state from  two pointlike sources 
% independently of the distance $|\vec{r}_1-\vec{r}_2|$ between them
% achieved the maximal SS (calculated over $R^3$).

Note that Plank constant $\hbar=1$ in our FM calibration, alike it's  
done in Relativistic QM;  in FM it only relates $x,p$ scales in 
our formalism and doesn't have any other meaning \cite {Fey}.
% We believe  that the results obtained here can
% have the general meaning for QM axiomatics considerations independently of
% FM hypothesis.
The proposed FM considers the nonrelativistic particle for which 
$x$ is the fuzzy coordinate, yet from the symmetry of phase space
one can choose any observable $Q$ as the fundamental fuzzy
coordinate and from this assumption to reconstruct FM formalism.
It can be especially important in  relativistic case where
 $\vec r$ can't be the proper observable \cite {Schiff}.
In addition, the linearity of state evolution becomes the important criteria
for the choice of consistent ansatz. For massive particle the minimal
solution truns out to be $4$-spinor, i.e. it responds to Dirac equation for
 spin-$\frac{1}{2}$ \cite {Schiff}.
 FM approach, in principle,
 can be extended on quite different physical systems. 
Here we regarded the fuzzy  phase space of single particle,
but such  phase space of any kind can be constructed.
 In particular, it can be Fock space
for the secondary quantization, in this case, the occupation numbers for 
particle's states $N_c(\vec{p})$ can be regarded as the fuzzy values.
  
%\section *{References}


\begin {thebibliography}{99}

\bibitem {Vax} S.Mayburov 2002 
%'Fuzzy Space-Time Geometry
%as Approach to Quantization',
 {\it Proc. of Quantum Foundations conference}
( Vaxjio Univ. Press, Vaxjo) p~232; hep-th 0210113 

\bibitem {May} S.Mayburov 2004
% 'Fuzzy Geometry of Space-time and
% Quantum Dynamics'
{\it Proc. Steklov Math. Inst.} {\bf 245} 154 

\bibitem{Aha} Y.Aharonov, T.Kaufherr 1984
%'Reference Frames of Quantum Mechanics'
 {\it Phys. Rev.}  {\bf D30} 368 
 
\bibitem {Dop} S.Doplicher, K.Fredenhagen, K.Roberts 1995
% 'Quantum  Space-Time at Plank Scale' 
 {\it Comm. Math. Phys.} {\bf 172} 187 

% \bibitem {Con} A.Connes 'Noncommutative Geometry' (Academic Press, 1994) 

% \bibitem {Rov} C.Rovelli, Class. Quant. Grav. 8, 317,(1991)
 
% \bibitem {Unr} W.G.Unruh, R.M.Walde Phys. Rev. D40, 2598, (1989) ,
%  H.Kitada Nuov. Cim. 109B , 281 (1995) , gr-qc/9708055

\bibitem {Ish} C.Isham 1994
%'Introduction into Canonical Gravity Quantization'
  {\it Canonical Gravity: from Classical to
Quantum } Eds. J.Ehlers, H.Friedrich , Lecture Notes in Phys. 433
( Springer, Berlin) p~11

\bibitem {Vol} V.S. Vladimirov, I.V. Volovich 1989
%P-adic Quantum Mechanics',
 {\it Comm. Math. Phys.}
 {\bf 659} 123 

% \bibitem {B} S.Baez et. al. Com. Math. Phys. 208, 787 (2000) 

\bibitem {Zad} L.Zadeh 1965
% 'Fuzzy Sets'
{\it Inform. and Control} {\bf8} 338
%  IEEE Trans., SMC-3, 28 (1973)

\bibitem {Got} H.Bandemer, S.Gottwald 1993
 {\it Einfurlung in Fuzzy-Methoden}
(Academie Verlag, Berlin) p~45


\bibitem {Zee} C.Zeeman 1961
% 'Topology of Brain and Visual Perception'
  {\it Topology of 3-manifolds } 
 (Prentice-Hall, New Jersy) p~240 

\bibitem {Dod} C.T.J. Dodson 1975
% 'Tangent Structures for Hazy Space'
%  Bull. London Math. Soc., 6, 191 (1974);\,    
 {\it J. London Math. Soc.} {\bf 2} 465 

\bibitem {Pos} T.Poston 1971 
% 'Fuzzy Geometry',
{\it Manifold} {\bf 10} 25 

\bibitem {Ali} T.Ali, G. Emch 1974
% 'Fuzzy Observables in Quantum Mechanics'
 {\it Journ. Math. Phys} {\bf 15} 176
% \, ibid., 18, 219, (1977)

\bibitem {Pyk} J.Pykaz 2000
% 'Lukasiewics Operations on Fuzzy Sets'
 {\it Found. Phys.}  {\bf 30} 1503 

% \bibitem {Ren} M.Requardt , S.Roy
% 'Quantum Space-Time as Statistical Geometry of Fuzzy Lumps'
%  Class. Quant. Grav  18 , 3039 -3058 (2001)


%  \bibitem {Mayb8} S.Mayburov, Proc. V Quantum communications and measurements
%  Conf.,Capri, 2000 (Cluwer,N-Y,2001), Quant-ph 0103161


\bibitem {Schiff} J.M.Jauch 1968 {\it Foundations of Quantum Mechanics}
 (Adison-Wesly, Reading) 


\bibitem {Vlad} V.S.Vladimirov  1971 {\it Equation of Mathematical Physics}
(Nauka, Moscow) 

\bibitem {Ber} F.Berezin, E.Shubin 1985
 {\it Schr$\ddot{o}$dinger Equation} (Moscow, Nauka) 

\bibitem {Schw} L.Schwartz 1961 {\it Methods Mathematique pour
 les Sciences Physique} (Hermann, Paris)

% \bibitem {Kol} A.Kolmogorov, 'Information Theory' (Moscow, Nauka, 1957)  

\bibitem {Fed} M.Fedoriuk 1987  {\it Asymptotics: Integrals and rows}
(Nauka, Moscow) 

 \bibitem {Ed} R.Edwards  1965 {\it Functional Analysis and Applications}
 (N-Y, McGrow-Hill)

\bibitem {Fey} R.Feynman,A.Hibbs 1965
 {\it Quantum Mechanics and Path Integrals}
(N-Y, Mcgrow-Hill)

\bibitem {Gold} H.D.Doebner, G.A.Goldin 1996 {\it Phys. Rev}
 { \bf A54}  3764  
%  \bibitem {Mayb9} S.Mayburov Proc. VII Marcel Grossman Conf.,
%  Jerusalem ,1997 (W.S.,Singapour,1998), hep-th 0007003.



 

\end {thebibliography}
\end {document}